\documentclass{aa}
\usepackage{natbib,amssymb}
\usepackage[pdftex]{graphicx,color}
\bibpunct{(}{)}{;}{a}{}{,}
\newcommand{\solm}{M$_{\odot}$}
\newcommand{\soll}{L$_{\odot}$}
\newcommand{\solr}{R$_{\odot}$}

\title{Cometary shaped sources at the Galactic Center}
\subtitle{Evidence for a wind from the central 0.2$\,$pc}
 \author{K. Mu\v{z}i\'{c}$^{1}$\thanks{Present address: Department of Astronomy and Astrophysics, 
University of Toronto, 50 St. George Street, Toronto ON M5S 3H4, Canada}, 
A. Eckart$^{1,2}$, R. Sch\"odel$^3$, R. Buchholz$^{1}$,
         M. Zamaninasab$^{1,2}$, and G. Witzel$^{1}$}
\institute{1) I. Physikalisches Institut, Universit\"at zu K\"oln,
           Z\"ulpicher Str. 77,
           50937 K\"oln, Germany\\
           2) Max-Planck-Institut f\"ur Radioastronomie,
           Auf dem H\"ugel 69,
	   53121 Bonn, Germany\\
           3) Instituto de Astrof\'isica de Andaluc\'ia CSIC,
              Glorieta de la Astronoma S/N, 18008 Granada, Spain\\
           \email{muzic@astro.utoronto.ca}
}

\date{Received  / Accepted }
\begin{document}
\abstract
{In 2007 we reported two cometary shaped sources in the
vicinity of Sgr$\,$A* (0.8" and 3.4" projected distance), named X7 and X3.
The symmetry axes of the two sources are aligned to within 5$^{\circ}$ in the
plane of the sky
and the tips of their bow-shocks point towards Sgr$\,$A*. Our measurements show that
the proper motion vectors of both features are pointing in directions more than
45$^{\circ}$ away from the line that connects them with Sgr$\,$ A*. This misalignment of the
 bow-shock
symmetry axes and their proper motion vectors, combined
with the high proper motion velocities of several 100$\,$km$\,$s$^{-1}$,
suggest that the bow-shocks must
be produced by an interaction with some external fast wind, possibly coming from
Sgr$\,$A*, or stars in its vicinity.}
{We have developed a bow-shock model
to fit the observed morphology and constrain the source of the external wind. }
{The result of our modeling gives the best solution for bow-shock standoff
distances for the two features, which allows us to estimate
the velocity of the external wind, making sure that all likely stellar types
of the bow-shock stars are considered.}
{
We show that neither of the two bow-shocks (one of which is 
clearly associated with a stellar source) can be produced by influence
of a stellar wind of a single mass-losing star in the central parsec. 
Instead, an outflow carrying a momentum comparable to the one contributed 
by the ensemble of the massive young stars, 
can drive shock velocities capable of producing the observed cometary features.
We argue that a collimated outflow arising perpendicular to the plane of the
clockwise rotating stars (CWS), can easily account for the two features and the mini-cavity.
However, the collective wind from the CWS has a scale of $>\,$10$''$.
The presence of a strong, mass-loaded outbound
wind at projected distances from Sgr$\,$A* of $<\,$1$''$ is in fact in
agreement with models that predict a highly inefficient accretion onto
the central black hole due to a strongly radius dependent accretion flow.
}
{}

\keywords{Galaxy: center -- Stars: mass-loss -- Infrared: stars -- Infrared: ISM}

\authorrunning{K. Mu\v{z}i\'{c} et. al.}
\titlerunning{Cometary shaped sources at the GC}

\maketitle


\section{Introduction}
\label{intro}
Analyses of stellar orbits in the central
arcsecond of the Milky Way have provided indisputable evidence that
the central object at the position of the radio source Sagittarius$\,$A* (Sgr$\,$A*)
is a supermassive black hole (SMBH; e.g. \citealt{eckart02,schoedel03,eisenhauer05,ghez05,ghez08,gillessen09}).
Sgr$\,$A* is located at the distance of $\sim\,$8$\,$kpc, with a mass of $\sim\,$4$\,\times\,$10$^6$\solm.

There is long-standing evidence for the existence of outflow(s) at the Galactic Center (GC).
The central parsec of the Milky Way harbors a cluster of
massive stars (e.g. \citealt{paum06}) that contributes $\sim\,$3$\times$10$^{-3}$\solm$\,$yr$^{-1}$ to the
center in
form of stellar winds \citep{najarro97}. However, less than 1$\%$ of this material is
actually accreted onto the SMBH \citep{baganoff03,bower03, marrone06} and thus has to be expelled out of the center.

The streamers of gas and dust in the central few parsecs
of the Galaxy show a bubble-like region of lower density
(called mini-cavity), located $\sim$3.5'' southwest of
Sgr$\,$A*. It
was first pointed out on cm-radio
maps by \citet{y-z90}. The strong Fe[III] line
emission seen toward that region \citep{eckart92,lutz93}
 is consistent with gas excited by a collision with a fast
($\geq$1000 km$\,$s$^{-1}$) wind from a source within the central few arcseconds
\citep{y-z93, y-z92}.

\begin{figure*}
\centering
 \resizebox{13cm}{!}{\includegraphics{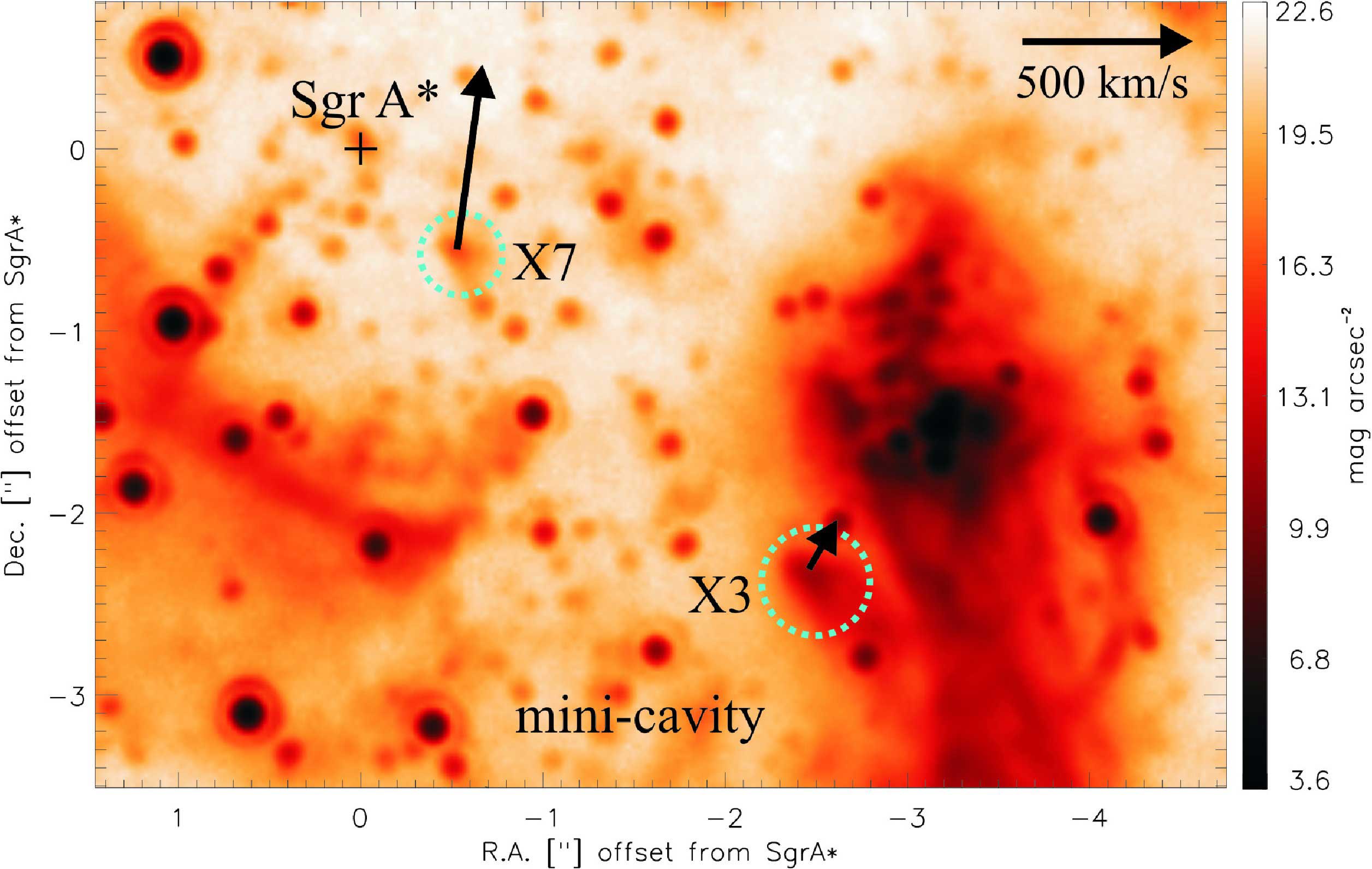}}
 \caption{NACO L'-band (3.8$\,\mu$m) image of the Galactic Center region showing projected positions
 of Sgr$\,$A* and the cometary-shaped features X3 and X7. The arrows show proper motions.
North is up, east is to the left.}
\label{SgrA*X3X7}
\end{figure*}

In \citet{muzic07} we presented NACO/VLT multi-epoch observations at 3.8$\,\mu$m (L'-band) which
allowed us to derive proper motions of narrow filamentary features associated with the gas and dust
streamers of the mini-spiral. 
Proper motions of several
bow-shock sources have also been reported.
The analysis has shown that the shape and the motion of the filaments is not consistent with a
purely Keplerian motion of the gas in the potential
of Sgr$\,$A* and that additional mechanisms must be responsible for their formation
and motion. We argued that the properties of the filaments are probably related to
an outflow from the disk of young mass-losing
stars around Sgr$\,$A*. In part, the outflow may originate from the
immediate vicinity of the black hole itself.

In \citet{muzic07} we report the existence
 of the two cometary-shaped features X3 and X7, located at
projected distances of 3.4'' and 0.8'' from Sgr$\,$A*, respectively.
The symmetry axes of the two bow shocks are almost aligned (within 5$^{\circ}$) and point
towards Sgr$\,$A* (see Fig.$\,$\ref{SgrA*X3X7}). 
At the same time, their proper motion vectors are not pointing in the
direction of the symmetry axes, as it would be expected if the bow-shock shape were produced
by a supersonic motion of a mass-losing star through a static interstellar medium. In general,
bow shock appearance can also be produced by an external supersonic wind
interacting with a slower wind from
a mass-losing star. 
If the velocity contrast of the two winds is large, one will observe
a bow shock pointing towards the source of the external wind.
If in addition the star moves in a direction different than the position
of the external wind, the bow shock will point in the direction
of the relative velocity between the star and the incident medium.

In this paper we present new proper motion measurements for
the two cometary-shaped sources X3 and X7 (section$\,$\ref{sec:pm}). In section$\,$\ref{model}
we explain details of the bow-shock modeling that was used to fit
those features (section$\,$\ref{results}). Possible
thee-dimensional positions of the sources are discussed in section$\,$\ref{sec:3D}.
In section \ref{discuss} we argue about the nature of the two sources and
the external wind held responsible for the observed arrangement.
Summary and conclusions are given in section$\,$\ref{summary}.


\section{Observations and Data Reduction}
\label{reduction}

The L' (3.8 $\mu$m)
images were taken with the NAOS/CONICA adaptive optics assisted imager/spectrometer
\citep{lenzen98,rousset98,brandner02}
at the UT4 (Yepun) at the ESO VLT.
The data set includes images from 7 epochs
(2002.66, 2003.36, 2004.32, 2005.36, 2006.41, 2007.25 and 2008.40)
with a resolution of $\sim$100$\,$mas and a pixel scale of 27$\,$mas/pixel.
Data reduction (bad pixel correction, sky subtraction, flat field correction)
and formation of final mosaics was performed using the DPUSER software for
astronomical image analysis (T. Ott; see also \citealt{eckart90}).

\section{Proper motions}
\label{sec:pm}
\begin{figure}
\centering
 \resizebox{9cm}{!}{\includegraphics{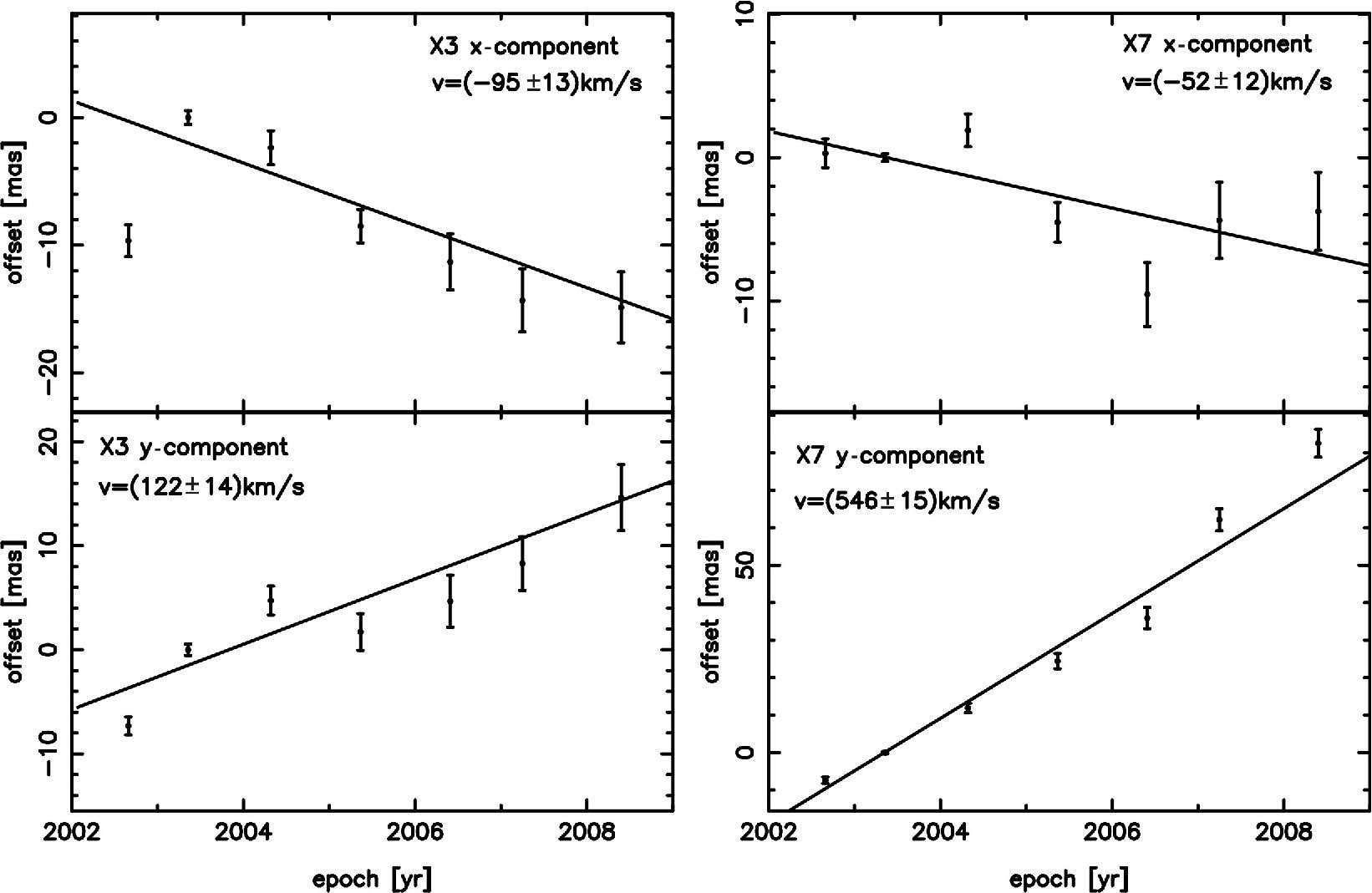}}
 \caption{L'-band proper motions of the two cometary features X3 and X7. 
 The error bars show the 1$\sigma$ uncertainty of each measurement. The $x$ and $y$ components 
 refer to RA and Dec, respectively.} 
\label{pmgraphs}
\end{figure}

In order to obtain proper motions of extended features we have used the
same method described in detail in \citet{muzic07}.
Our measurements are shown in Fig.$\,$\ref{pmgraphs}.
 The distance to the GC is assumed to be 8$\,$kpc.
The results obtained from the L'-band data between 2002 and 2008, are as follows.
\textbf{X7:}
v$_{\alpha}$=(-52$\,\pm\,$12)$\,$km$\,$s$^{-1}$,
v$_{\delta}$=(546$\,\pm\,$15)$\,$km$\,$s$^{-1}$.
\textbf{X3:}
v$_{\alpha}$=(-95$\,\pm\,$13)$\,$km$\,$s$^{-1}$,
v$_{\delta}$=(122$\,\pm\,$14)$\,$km$\,$s$^{-1}$.
The uncertainties are 1$\,\sigma$ uncertainties of the weighted linear fit to the
positions vs. time.
The proper motion velocity vectors of both sources are
oriented in the northwest quadrant (see Fig.$\,$\ref{SgrA*X3X7}).
We note here that the results are in agreement with results given in Table$\,$1. of
\citet{muzic07}, but
there is  an error in their Fig.$\,$5., where X3 is plotted as moving towards north-east.

The L-band feature X7 is coincident (in projection) with a K-band source S$\,$50 \citep{gillessen09}. The reported proper motion of S$\,$50 is v$_{\alpha}$=(-108$\,\pm\,$5)$\,$km$\,$s$^{-1}$, and
v$_{\delta}$=(361$\,\pm\,$7)$\,$km$\,$s$^{-1}$. While the orientations of the two proper motions are
in agreement, there is a significant difference in proper motion magnitudes.
This, however, might give us an idea
about the systematic uncertainties of the proper motions derived from the L-band extended sources.
It seems reasonable to argue that we are not dealing with an accidental superposition of 
a stellar source with a dust blob along the line of sight, but that the L-band feature
is indeed associated with the stellar source at the same position.


\section{Model}
\label{model}
\subsection{Bow-shock shape}
\label{shape}
\subsubsection{Analytic model}
\label{analytic}

\begin{figure}
\centering
 \resizebox{9cm}{!}{\includegraphics{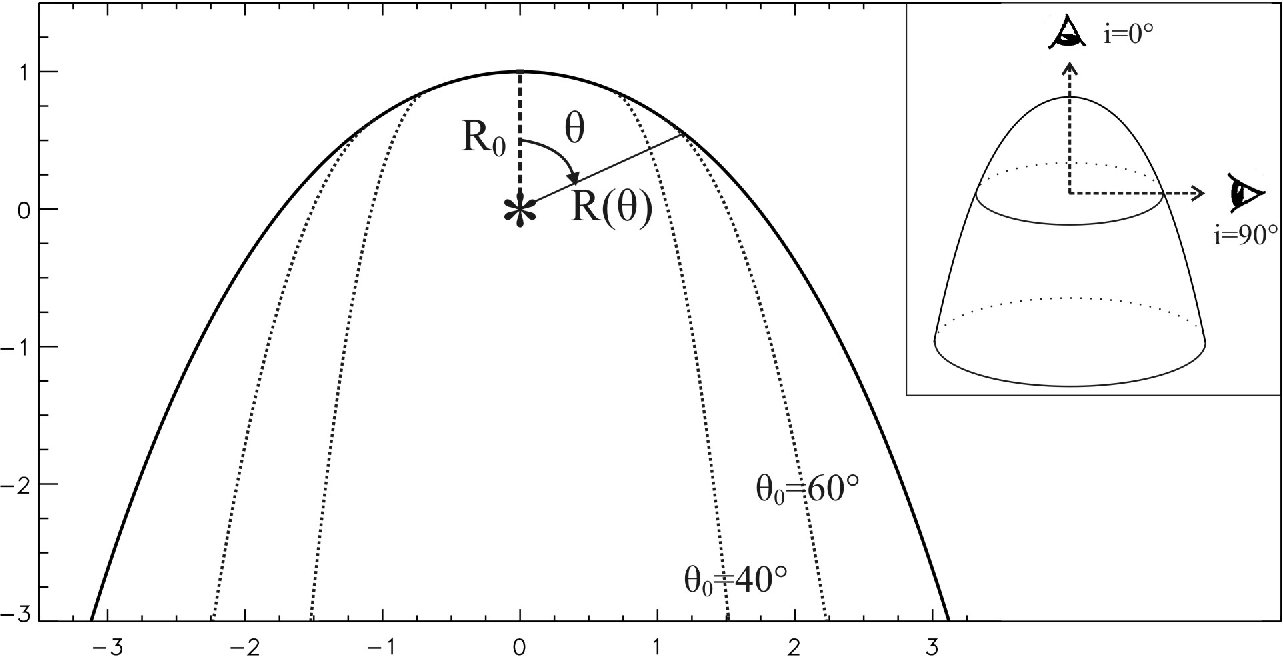}}
 \caption{The two-dimensional bow-shock shape.
Full line shows the analytic model from \citet[$\,$see Eq.$\,$\ref{Req}]{wilkin96PhD}.
 Dotted lines show narrow solutions \citep{zhang&zheng97}, for two
 collimation angles $\theta_0$=40$^\circ$ and $\theta_0$=60$^\circ$.
The axes are in units of $R_0$; the
 star is at the center of the coordinate system.\newline
\textit{Inset:} 3D bow-shock sketch, explaining the conventions for the inclination angle used in this work.
 $i$=0$^\circ$ when the bow-shock symmetry axis is parallel to the line of sight (i.e
we are looking directly at the apex). $i$=90$^\circ$ if the bow-shock symmetry axis is
perpendicular to the line of sight. For a bow-shock inclined away from the observer (the observer looks at the
envelope, but cannot see the apex), 90$^\circ<i<$180$^\circ$.}
\label{modelsketch}
\end{figure}

\citet{wilkin96PhD} derived a fully analytic solution for the shape of
a bow shock produced by a star moving through the interstellar
 medium at supersonic velocity (see Fig. \ref{modelsketch}).
The two-dimensional shell shape is given by:
\begin{equation}
 R(\theta)=R_0 \csc \theta \sqrt{3(1-\theta \cot\theta)},
 \label{Req}
\end{equation}
where $\theta$ is the polar angle from the axis of symmetry, as seen by the star at the coordinate origin.
$R_0$ is the so-called standoff distance obtained by balancing the ram pressures
of the stellar wind and ambient medium at $\theta$=0 and is given by:
 \begin{equation}
R_0=\sqrt{\frac{\dot{m}_wv_{w}}{4\pi\rho_av_{a}^2}}.
\label{R0eq}
  \end{equation}
Here $\dot{m}_w$ is the stellar mass-loss rate, $v_{w}$ the terminal velocity of the stellar wind,
$\rho_a$ the ambient medium mass density, and $v_{a}$ is the velocity at which the star moves through
the medium (i.e. the relative velocity of the ambient medium and the star velocity
in case that the ambient medium is not stationary).

We assume that the shell has a thickness as shown in Fig$\,$3b in \citet{maclow91}. 
Then we rotate the two-dimensional
shape around its axis of symmetry in order to obtain the three-dimensional shell.

\subsubsection{Narrow solution}
\label{narrow}
The model by \citet{wilkin96PhD} incorporates the \textsl{instantaneous cooling approximation}: the
interaction between the stellar wind and the ambient medium takes place in an infinitely
thin layer, the two flows are fully mixed and immediately cooled. In this case $R_0$ directly gives
the distance of the star to the apex of the bow shock. However,
\citet{comeron98} note that this might not be true if cooling of the shocked stellar wind is inefficient. In
this case it is expected that the bow shock would be located at a distance somewhat greater than $R_0$
\citep{comeron98, raga97, povich08}. For some GC sources,
\citet{tanner05} indeed reported that a better fit can be obtained
if the apex of the bow shock is shifted away from the star. We have observed the same behavior
for several other Galactic Center bow shocks. 

\citet{zhang&zheng97} have investigated the case where the wind ejection from the star is not
necessarily isotropic; it is rather confined in a cone of solid angle $\Omega=2\pi(1-\cos\theta_0)$, where
2$\theta_0$ is the opening angle in which the matter is ejected. The standoff distance is then given by:
 \begin{equation}
R_0=\sqrt{\frac{\dot{m}_wv_{w}}{\Omega\rho_av_{a}^2}}.
\label{R0eqnarrow}
  \end{equation}
As can be seen in Fig.$\,$\ref{modelsketch}, with $\theta_0$ getting smaller, the bow shock
profile will have a narrower shape. For $\theta_0\rightarrow 3\pi/4$ the shape is very similar
to the one given by Eq.$\,$\ref{Req}.
Although this model has been developed for the case of a collimated stellar wind,
we do not assume such a collimation, since it would imply that the collimation
of the wind is in the direction of motion of the star, relative to the ambient medium. There is
no a priori reason for the stellar velocity and wind collimation direction to be linked.
We introduce $\theta_0$ as an additional factor that allows one to control the width 
of the bow shock outline, while causing 
the apex of the bow shock to be slightly displaced away from the star.

\subsection{Generating emission maps}
\label{emission}

To generate a simulated observation, we start by illuminating the shell by the star placed at the
origin of the coordinate system and then calculate the emission along each ray intersecting the shell.
The shell is inclined to the line of sight by the angle $i$, and also rotated
 by a position angle PA in the plane parallel to the plane of the sky.
Inset in Fig.$\,$\ref{modelsketch} explains the convention used for the inclination angle. PA is measured 
East of North.

Each parcel of the shell is assigned the optical depth $\tau$, calculated as:
 \begin{equation}
\tau(\lambda)= \tau_{abs}(\lambda)+\tau_{sca}(\lambda)= L\int_{a_{-}}^{a_{+}}n_d(a)C_{ext}(a,\lambda)da.
  \end{equation}
We assume a graphite+silicate mixture with a power-law grain size distribution $n_d(a)$ \citep{MRN77},
 where $a$ is the grain size.
Dust extinction coefficient is $C_{ext}=\pi a^2(Q_{abs}+Q_{sca})$, where $Q_{abs}$ and $Q_{sca}$ are
dust absorption and scattering efficiencies, respectively \citep{laor&draine93}
\footnote{available at http://www.astro.princeton.edu/\~{}draine/}. $L$ is the length of the
shell parcel along the line of sight.

The source X7 is polarized \citep[see][]{muzic07}, suggesting that
scattering of the stellar emission by dust particles in the bow-shock envelope
is probably important. In order to account for scattering, we proceed in the following way:
Emission of each parcel of the shell has
contributions both from scattering and thermal emission: $L$=$L_{sca}$+$L_{th}$.
$L_{th}\propto$$\,B(T_d)$$(1-e^{-\tau_{abs}})\epsilon_{th}$ and
$L_{sca}\propto$d$^{-2}$$\epsilon_{sca}P(\theta_{sca})$$e^{-\tau_{sca}}$, where
$d$ is the distance from the star, 
$B(T_d)$ is the black body emission at the dust temperature, integrated 
over grain sizes, and over wavelengths in our observing band.
 $\epsilon_{th}$ and $\epsilon_{sca}$ are
thermal emission and scattering efficiencies, respectively. 
Dust temperature at a distance $d$ (in parsecs) 
from the star can be calculated as
$T_d\,=\,$27$\,a$$_{\mu m}^{-1/6}$$\,L$$_{*,38}^{1/6}$$\,d$$_{pc}^{-1/3}$$\,$K \citep{vanburen88, kruegelbook},
 where $a_{\mu m}$ is dust grain size in $\mu m$, and $L_{*,38}$ is stellar luminosity in 10$^{38}$ergs$\,$s$^{-1}$.
$P(\theta_{sca})$
is the (normalized) scattering function that controls the amount of forward scattering, and
is given by:
 \begin{equation}
P(\theta_{sca})=\frac{1-g^2}{1+g^2-2 g cos(\theta_{sca})}.
 \end{equation}
For $g=0$, scattering is isotropic and $P$ does not depend on the scattering angle $\theta_{sca}$.
$g=1$ means full forward scattering. We have varied $g$ between 0 and 0.8, in steps of 0.2.
Since $\epsilon_{th}$ and $\epsilon_{sca}$ are in general not known, we have chosen to
test three different cases: ($\epsilon_{th}$, $\epsilon_{sca}$) = (1.0, 0.0), (0.0, 1.0), and (0.5, 0.5), 
i.e. thermal emission only, scattering only, and equal contributions from both processes.
The shape of the thermal emission depends on the dust temperature distribution: we investigate central
stars with luminosities (10$^2$, 10$^3$, 10$^4$)\soll.

Finally, the resulting projection of the 3D-geometry onto the plane
of the sky is rebinned to the pixel-scale of the NACO images and smoothed
with a gaussian PSF having FWHM equivalent to the angular
resolution of our images.

\section{Results}
\label{results}

\subsection{X7}
\label{X7}

\begin{figure*}
\centering
 \resizebox{18cm}{!}{\includegraphics{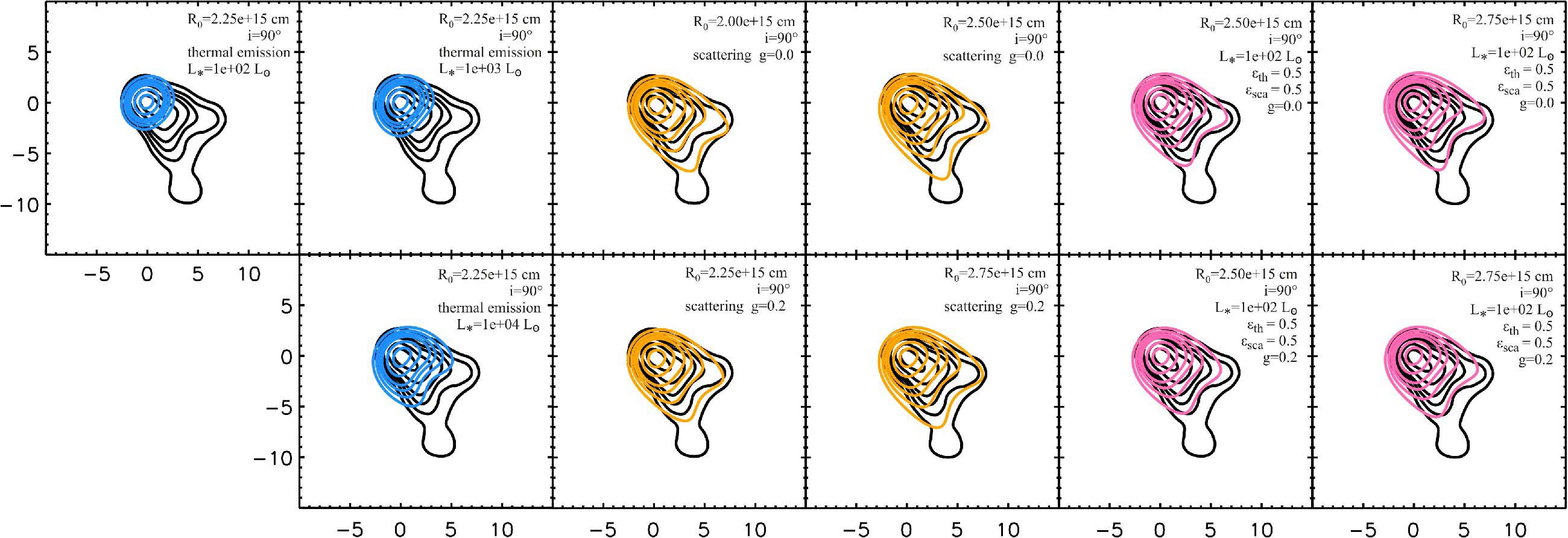}}
 \caption{Results of our modeling for the feature X7 (black contours),
 observed in the epoch 2003.36 (the image is previously deconvolved and beam-restored to 
the nominal resolution of our L'-band images). 
Color contours represent the bow-shock model projected onto
the plane of the sky.
PA$\,$=$\,$50$^{\circ}$ (E of N), and the shape is given by the analytic solution (section \ref{analytic}).
We show purely thermal emission in blue (left), pure scattering in orange (middle),
 and equal contributions from both processes in purple (right).
For each set of parameters, we show the results for
 two different values of $R_0$, in order to depict how change in the bow-shock size affects
the fit.
 Contour levels are at (20, 30, 40, 50, 70, 90)$\%$
of the maximum. North is up, east is to the left. The axes are labeled in pixel offsets
from the brightest point in the image; 1 pixel is equivalent to 0.027$\,$mas.
}
\label{X7fit}
\end{figure*}

Fig.$\,$\ref{X7fit} shows the
best-fit results of our bow-shock modeling for the feature X7.
We have tested the three different combinations of ($\epsilon_{th}$, $\epsilon_{sca}$),
and five values of $g$=(0.0, 0.2, 0.4, 0.6, 0.8). In Fig.$\,$\ref{X7fit}, however,
we show only those values that result in good fits, except for the left-most
panels (blue contours), which we show to illustrate the behaviour of purely thermal emission.
Thermal emission is dominant in the vicinity of the star, but cannot fit the
extended tail that we observe in X7, unless the central star is much brighter than
the cases investigated here. This, however, is not plausible since X7 is very faint 
in the K-band (see discussion
below). The tail of the bow-shock is much better described by scattering.
For ($\epsilon_{th}$, $\epsilon_{sca}$)=(0.5, 0.5), we show only the solution for 
L$_{*}$=10$^2$\soll. We choose to do so because the difference in resulting contours 
for the three stellar luminosities is small, and gives the same solution for
the physically most interesting parameter, $R_{0}$.
In cases when scattering is important ($\epsilon_{sca}\geq$0.5), more
forward scattering (large g) results in a more compact model, thus not fitting well the outer
contours. There is no significant influence of the parameters on the inner contours, due to
relatively small size of the feature, and smoothing. These differences can be better observed in case of X3.

For each set of parameters ($\epsilon_{th}$, $\epsilon_{sca}$, g), we have tested different
values of $R_0$. By changing $R_0$, the model preserves the same shape of the 
contours, but is as a whole expanded or shrunken. Therefore in Fig.$\,$\ref{X7fit} we 
plot two values of $R_0$ for each set of parameters. We chose values in the way
that shows how the change in $R_0$ affects the fit. By changing its value by larger
amount, the fit becomes inadequate. 

The best solutions for different sets of parameters 
are obtained for $R_0\approx\,$2.5$\cdot$10$^{15}$cm, and 
PA$\,$=$\,$50$^\circ$, measured east of north.
Note that in this case the best results are
obtained using the simple analytic two-dimensional solution (eq.$\,$\ref{Req}).
Both X7 and X3 have an unusually narrow appearance, and therefore are best
fitted with inclination angles close to 90$^\circ$.

X7 coincides with a point source at shorter wavelengths (see discussion on
proper motions in section~\ref{sec:pm}).
Photometric measurements give H=18.9$\pm$0.1 and K=16.9$\pm$0.1 \citep{schoedel10}.
For the local extinction at the position of X7 we assume A$_K$=2.5 \citep{schoedel10}.

In section \ref{X7discuss} we discuss possible stellar types and implications this has on the external
wind parameters.

\subsection{X3}
\label{X3}

\begin{figure*}
\centering
\resizebox{12cm}{!}{\includegraphics{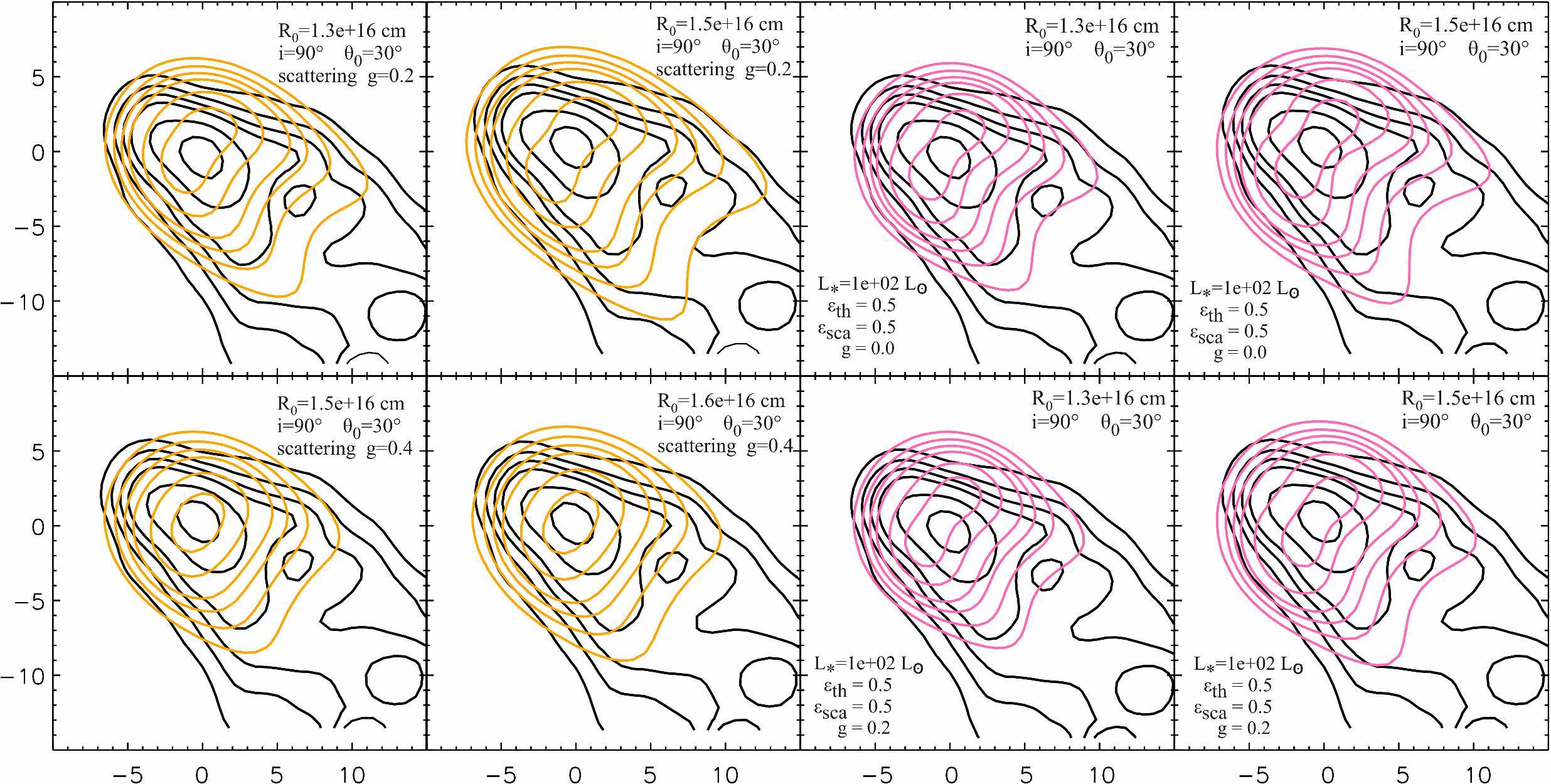}}
 \caption{Best fit results of the modeling for the feature X3 (black contours),
 observed in the epoch 2003.36 (the image is previously deconvolved and beam-restored 
to the nominal resolution of our L'-band images). 
Color contours represent the bow-shock model projected onto
the plane of the sky.
PA$\,$=$\,$55$^{\circ}$ (E of N) and $\theta_0$=30$^{\circ}$ in all cases.
We show the pure scattering case in orange (left),
 and equal contributions from scattering and thermal emission in purple (right).
For each set of parameters, we show the results for two different values of $R_0$, 
in order to depict how change in the bow-shock size affects the fit.
Contour levels are at (20, 30, 40, 50, 70, 90)$\%$
of the maximum. North is up, east is to the left. The axes are labeled in pixel offsets
from the brightest point in the image; 1 pixel is equivalent to 0.027$\,$mas.
}
\label{X3fit}
\end{figure*}

Fig.$\,$\ref{X3fit} shows the
best-fit results of our bow-shock modeling for the feature X3.
This feature is very elongated and a satisfactory fit cannot be obtained using
the analytic 2D solution. It requires a narrow model (see Section \ref{narrow}), with
small opening angles $\theta_0$.
As in the case of X7, the outer contours are represented better in models with lower $g$, while
larger $g$ values result in more compact inner contours.
Here it is even more evident that thermal emission gives a too
compact model. The elongated tail of X3 can only be well fitted with models
that include scattering.
Therefore we show only these solutions, in pairs of two different values of $R_0$. 
The best fit
solutions give $R_0\approx\,$1.5$\cdot$10$^{16}$cm, with $\theta_0$=30$^{\circ}$,
$i\,$=$\,$90$^\circ$, and
PA$\,$=$\,$55$^\circ$.

In contrast to X7, there is no detectable point source at the
position of X3 in our K$_S$-band images. Local extinction at the position
of X3 is  A$_K\approx$2.7 \citep{schoedel10}.
In section \ref{X3discuss} we discuss the possible nature of this source and the
implications this has on the external
wind parameters.

\section{3-dimensional arrangement}
\label{sec:3D}
\begin{figure*}
\centering
 \resizebox{18cm}{!}{\includegraphics{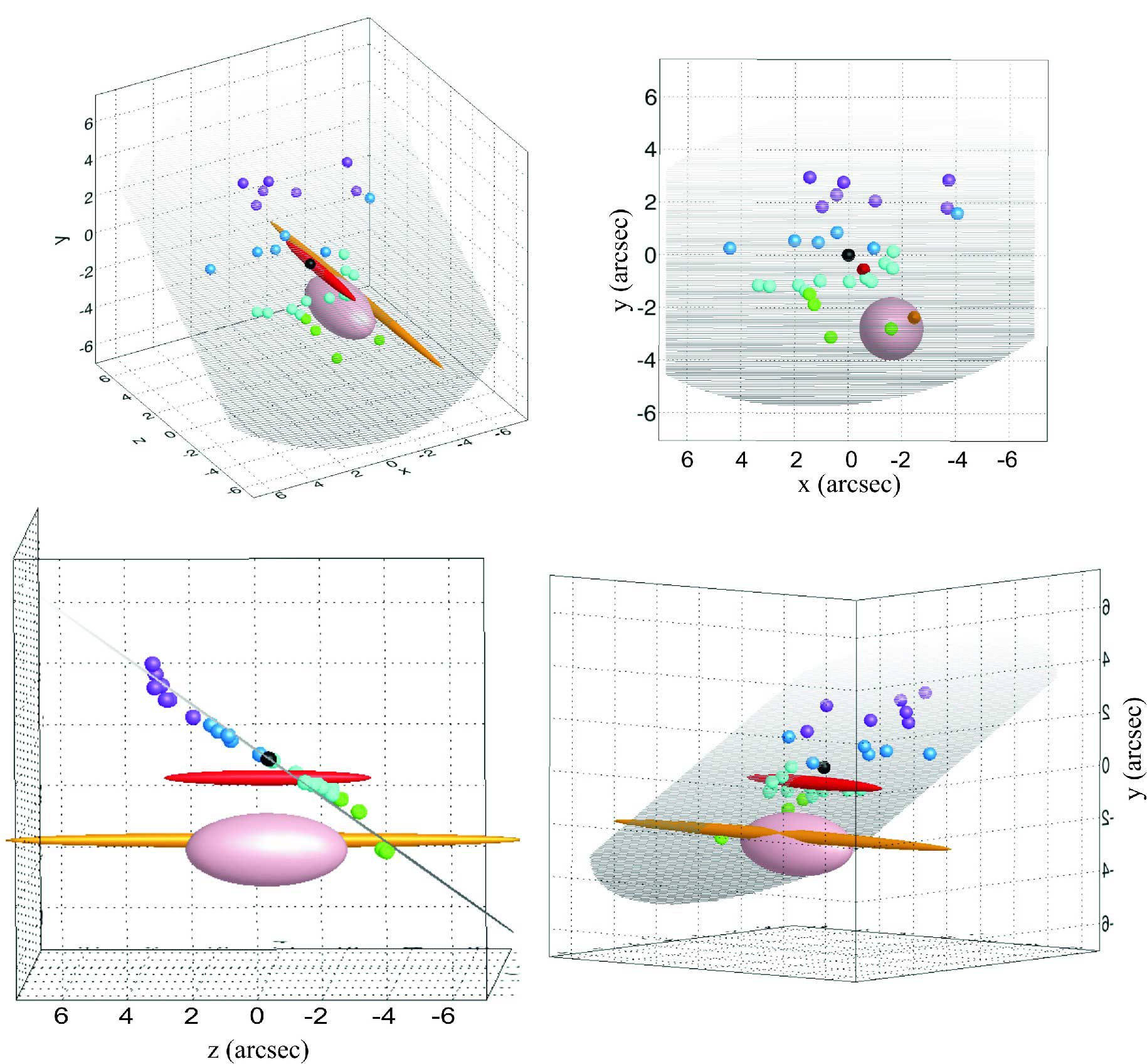}}
 \caption{Three-dimensional view of some of the Galactic Center features. The axes show offsets
from Sgr$\,$A* (black sphere) in arcseconds. On the z-axis, positive means further away from the
observer than Sgr$\,$A*.
The shaded area represent the CWS disk and
the colored spheres stars belonging to it; color scheme reflects the distance from
the observer, with green being closest and violet farthest away from us.
The bow shock sources are shown in red (X7) and orange (X3). Elongation along
the z-axis reflects the uncertainty in the position of the two sources along
the line of sight (see text). The pink spheroid represents the mini-cavity: in projection
we plot it as a circle with radius $r\,\sim\,$1.2'' and the elongation
along the z-axis reflects the range of possible positions, as given by \citet{paum04} and
\citet{zhao09}.
We show the same setup from four different angles; the top right panel shows the projection
onto the plane of the sky.}
\label{GC3D}
\end{figure*}

Fig.$\,$\ref{GC3D} shows a 3D reconstruction of some of the features found
in the central parsec of the Galaxy. The shaded area represents the disk of
clockwise-rotating stars (CWS; \citealt{paum06}; \citealt{beloborodov06}; \citealt{lu09}), and the
colored spheres are the stellar members. The positions of the
stars and the disk parameters ($n_x, n_y, n_z$)=(-0.12, -0.79, 0.6) are from \citet{paum06}.
The stars are represented by different colors according to their distance
from the observer (green is closer and violet is further away from us).

\citet{eckart02} show how a three-dimensional separation $r$ from the
center can be estimated by comparing the proper motion (V$_{PM}$) of a star to the
three-dimensional velocity dispersion $\sigma$.
The probability that a star at the position $r$
has a proper motion larger than V$_{PM}$ can be calculated via
\begin{equation}
 P(V>V_{PM},r) = 1-\frac{1}{\sigma^2}\int_0^{V_{PM}}v~exp\Big(\frac{-v^2}{2\sigma^2}\Big)dv.
\label{eq:P}
\end{equation}
The velocity dispersion $\sigma(r)$, and therefore also the probability P(V$>$V$_{PM}$,r),
decrease with increasing radii r. For a fast-moving star it becomes
increasingly unlikely that they belong to statistical samples at correspondingly larger radii. 
Therefore P(V$>$V$_{PM}$,r)
can be interpreted as a measure of how likely it is that the star belongs
to a sample of stars at radius $r$ or larger. Using the projected position $R$ instead
of the 3D value $r$, one can calculate $P(V>V_{PM},R)$ 
as an upper limit of the probability $P(V>V_{PM},r)$.
For a given radius $r$ we can calculate the velocity dispersion as $\sigma^2=GM_{BH}/r$, where
$M_{BH}$ is the mass of Sgr$\,$A*. This gives a good estimate of $\sigma$ within the central
parsec where the dynamics are dominated by the gravitational potential of Sgr$\,$A*.
Using this method, we estimate the line-of-sight position of X7 to be between $z=-3.2''$ and
$z=3.2''$, with P=67$\%$.
The elongation of the red object in Fig.$\,$\ref{GC3D} reflects this range of possible z-positions
for X7. The 3D position of the bow-shock X3 cannot be constrained that well due to its
lower proper motion and larger distance from the center. Therefore the orange object representing
X3 is stretched across the entire z-axis.

The mini-cavity is shown in pink: in projection we approximate it with a circle of $\sim$1.2'' radius.
\citet{paum04} argue that the mini-cavity is a part of the Northern Arm of the Mini-spiral. In
their reconstruction of the streamer 3D morphology, the mini-cavity should be exactly in the plane
parallel to the plane of the sky and containing
Sgr$\,$A*, or within $\sim$2.5'' from it. This is in excellent agreement with orbital
models of \citet{zhao09}.
The $\pm$2.5'' uncertainty of the line-of-sight position of the mini-cavity is
depicted as an elongation of the pink object in Fig.$\,$\ref{GC3D} 
along the z-axis.

\section{Discussion}
\label{discuss}
All the interesting physics that can result from the modeling
is contained in the equation for the bow shock standoff distance $R_0$ (eq.$\,$\ref{R0eqnarrow}).
The main drawback is a lack of knowledge about the stellar types (i.e. stellar wind parameters)
of the two stars. The L'-band images are dominated by thermal emission of dust. Both X3 and X7 are bright
at 3.8$\,\mu$m, but extremely faint at shorter wavelengths. Both features are apparently not embedded in the
mini-spiral material and therefore we expect them to be intrinsically dusty.

If there was no influence of any external wind, the bow shock in our images would
point in the direction of the proper motion. We understand that the $v_a$ in equations
(\ref{R0eq}) and (\ref{R0eqnarrow}) is the relative velocity
between the stellar velocity vector and the external wind velocity vector.
Since we do not have full three-dimensional information about the stellar velocity, but only
proper motions, the values of $v_a$ that we calculate in the following  represent lower limits for
the real external wind velocity.
For the orientation of the projected vector $v_a$ we assume
the position angle PA of the bow-shock symmetry axis (Figs.$\,$\ref{X7fit} and \ref{X3fit}).
By estimating $v_a$ we can therefore determine the projected position of the external wind
source responsible for the observed bow shock morphology.

\subsection{Nature of the source at the position of X7}
\label{X7discuss}
In the following we discuss the possible nature of the X7 star, based
on its apparent brightness in the K-band (16.9$\pm$0.1), combined with the distance modulus of 14.5
and extinction of $A_K\approx$2.5. We discuss only stellar types that are in agreement
with the K-band photometry.
 
For the number density of the ambient medium we assume $n\,$=$\,$26$\,$cm$^{-3}$ \citep{baganoff03}.
This value is reasonable since X7 is not moving through the denser mini-spiral material.\\

\noindent\textbf{(1) Late B-type main sequence star} (B7-8V).
For O and B galactic stars the following relation holds:
\begin{equation}
 log(\dot{m}_w v_{w}R_{\star}^{0.5})=-1.37+2.07log(L_{\star}/10^6),
\label{OB:eq}
\end{equation}
where $\dot{m}_w$ is in \solm$\,$yr$^{-1}$, $v_w$ in km$\,$s$^{-1}$, $R_{\star}$ in \solr\, and $L_{\star}$
in \soll\, \citep{LamersBook}.
For a typical B7V star with R$_{\star}\sim\,$4$\,$\solr\, and L$_{\star}\sim$10$^{2.5}$\soll\,we then have
$\dot{m}_wv_{w}\sim$10$^{-9}$\solm$\,$yr$^{-1}\,$km$\,$s$^{-1}$.
This kind of weak wind would require  $v_a$ of only 15$\,$km$\,$s$^{-1}$, a negligible velocity
when compared to the proper motion of X7.
Also, B-type main sequence stars are dust-free objects and thus probably
not good candidates to produce features like X7.\\

\noindent\textbf{(2) Herbig Ae/Be (HAe/Be) star.}
This class of intermediate-mass pre-main sequence objects is
characterized by strong wind activity and infrared excess. Line emission of
HAe/Be stars often shows prominent P-Cygni profiles, indicating powerful
winds with mass loss rates varying from  10$^{-8}$ to several times 10$^{-6}$\solm$\,$yr$^{-1}$, and
wind velocities of several hundred km$\,$s$^{-1}$ \citep[e.g.][]{benedettini98, bouret98, nisini95}.
\citet{nisini95} show that there is a correlation between mass
loss rate and bolometric luminosity of HAe/Be stars. A star with $L_{\star}$=10$^{2.5}$\soll\, would
have $\dot{m}_w\sim\,$10$^{-7}$\solm$\,$yr$^{-1}$.
Combined
with $v_w\sim$500$\,$km$\,$s$^{-1}$, this leads to
$v_a\sim\,$3000$\,$km$\,$s$^{-1}$. The external wind is then blowing with the velocity
$v_{ext}\sim$2800$\,$km$\,$s$^{-1}$ from the direction $\sim$60$^{\circ}$ (E of N).
We have to note that the
existence of the pre-main sequence stars at the GC is not established. It is not yet clear
how the young (few Myr old) population in the central parsec has been formed, since
the tidal field of Sgr$\,$A* prevents ''normal'' star formation via
cloud collapse \citep[e.g.][]{nayakshin06b, p-z06}. However, the
possibility of YSO presence in the central parsec was discussed by \citet{eckart04a} and \citet{muzic08}.
\\

\noindent\textbf{(3) Central stars of planetary nebulae (CSPN)}.
When all the hydrogen has been exhausted and the helium ignites in the core of
 a low/intermediate mass star (1$\,$-$\,$8\solm), the star reaches the asymptotic giant branch (AGB).
We note here that a typical AGB star at the GC is $\sim$4 magnitudes brighter in the K-band than X7.
AGB stars experience high mass-loss rates which remove most of
the stellar envelope, leaving behind a stellar core. The star has now entered the
post-AGB and subsequently planetary nebula (PN) phase. During this phase
the central star remains at constant luminosity, while
T$_{eff}$ rises, i.e. the star moves towards the left side of the HR diagram in a
horizontal line. CSPN are characterized by small mass losses, but very fast
stellar winds.
A typical CSPN with  $\dot{m}_w\sim\,$10$^{-8}$\solm$\,$yr$^{-1}$,
$v_w\sim$3000$\,$km$\,$s$^{-1}$, and R$\sim$\solr\, \citep{LamersBook}, will have
L=3$\times$10$^4$\soll\, and therefore $m_K$=17$\pm$1
for logT$_{eff}$$\sim$(4.6-4.8). The evolution of a star from the AGB phase towards
the final white dwarf stage is rapid \citep[e.g.][]{bloecker95}, and the star
will remain at the required brightness for less than 1000$\,$yr before it
becomes too faint. Adopting the typical CSPN parameters,
we obtain $v_a\sim\,$2300$\,$km$\,$s$^{-1}$, and
$v_{ext}\sim$2100$\,$km$\,$s$^{-1}$ from the direction $\sim$62$^{\circ}$ (E of N).\\

\noindent\textbf{(4) Low-luminosity Wolf-Rayet (WR) star; [WC]-type star.}
Population I WR stars are massive stars in a late stage of evolution.
With typical luminosities between 3$\times$10$^4$ and 10$^6$ \soll, they
are several magnitudes brighter than X7, unless the local extinction
at the position of X7 is much higher than the estimated value of A$_K$=2.5.
A class of CSPNe identified as WR stars share the origin with other PNe stars, but
are spectroscopically classified as WR stars because emission-line spectra
for both types of objects come from expanding H-poor stellar winds. WR stars in PNe almost
exclusively appear to be of the WC type \citep[e.g.][]{gorny95}, and are labeled as [WC].
About 15$\%$ of all the observed CSPNe appear to be [WC] \citep{acker03}.
PNe stars with WR stars in their centers have, in general, larger infrared excess than normal
CSPNe \citep{kwokPNe}, and more powerful winds. \citet{crowther08} reviewed
 the properties
of [WC] stars. Typically, winds are characterized by
$\dot{m}_w\sim\,$10$^{-7}$-10$^{-6}$\solm$\,$yr$^{-1}$ and $v_w\sim$200$\,$-2000$\,$km$\,$s$^{-1}$.
Adopting the lower limit values for $\dot{m}_w$ and $v_w$, we obtain
$v_a\sim\,$1900$\,$km$\,$s$^{-1}$,
$v_{ext}\sim$1700$\,$km$\,$s$^{-1}$ from the direction $\sim$65$^{\circ}$ (E of N).
For the average values $\dot{m}_w\sim\,$10$^{-6.5}$ and $v_w\sim$1000$\,$km$\,$s$^{-1}$,
we have $v_a\sim\,$7700$\,$km$\,$s$^{-1}$,
$v_{ext}\sim$7400$\,$km$\,$s$^{-1}$ from the direction $\sim$55$^{\circ}$ (E of N).

\subsection{Nature of the source at the position of X3}
\label{X3discuss}
In contrast to X7, there is no detectable point source at the
position of X3 in our K$_S$-band images. Due to crowding, individual stars
of K$_S\approx$18 are difficult to detect within the central parsec
and therefore we define this value as an upper limit on the brightness of X3 in the K-band.
Here we discuss the possible nature of the star at the position of X3, and
give possible solutions for the external wind direction and velocity.

\noindent\textbf{(1) Main sequence star.}
If on the main sequence, a star with K$\geq$18
would be of type A0 or later. The mass-loss rate of
low mass stars during the core H-burning phase is
very small, of the order of 10$^{-12}$ - 10$^{-10}$\solm$\,$yr$^{-1}$
\citep{LamersBook}.
In this case, any ISM streaming with the velocity of the order
10$\,$km$\,$s$^{-1}$ could produce the bow shock with
the standoff distance of X3. As in the case of X7, a
main sequence star cannot be a source of a dust-rich envelope
observed in the L'-band, and therefore we can rule out the main sequence
nature of X3.

\noindent\textbf{(2) CSPN or [WC]-type star.}
As a CSPN evolves on the horizontal track in the HR diagram, it maintains
the same luminosity while increasing the effective temperature. The peak emission
of the stellar black body shifts bluewards and the star becomes fainter in
the infrared. A typical CSPN or [WC]-star discussed above in case of X7, would
have $m_K>$18 for logT$_{eff}$$>$4.8.
Adopting the typical CSPN parameters,
we obtain $v_a\sim\,$1520$\,$km$\,$s$^{-1}$, and
$v_{ext}\sim$1540$\,$km$\,$s$^{-1}$ from the direction $\sim$61$^{\circ}$ (E of N).
In case of [WR]-star, using lower limit values as above, we obtain
 $v_a\sim\,$1240$\,$km$\,$s$^{-1}$, and $v_{ext}\sim$1260$\,$km$\,$s$^{-1}$ from
the direction $\sim$62$^{\circ}$ (E of N).

\noindent\textbf{(3) Dust blob.}
Since with the current sensitivity of our observations we do not
detect a point source in the K-band, we cannot rule out the possibility that
X3 is a dust blob ablated by a wind from the direction
of Sgr$\,$A*,
rather than a stellar source.
This scenario could explain the observed elongated tail.
The proper motion suggests that the feature moves along with the rest of the
material in the IRS13 region.
 In this case modeling
of the feature as bow shock is superfluous.

\subsection{Nature of the external wind}

The first question that
we have to ask is about a wind source
that can drive a shock of a certain velocity over a distance of few tenths of parsec.
The argument of the ram pressure of such a wind at the given
distance $d$ leads to
\begin{equation}
\label{Eq:wind}
 \rho_av_s^2=\frac{\dot{M}_wv_w}{4\pi d^2}.
\end{equation}
$\dot{M}_w$ and $v_w$ are the mass loss-rate and velocity
of the wind. The assumption behind Eq.$\,$\ref{Eq:wind} is
that the wind emanates isotropically from a point source.
Let us consider  the case of a single star
with the isotropic wind and typical wind parameters of the GC mass-losing stars:
$\dot{M}_w$=10$^{-5}$\solm$\,$yr$^{-1}$ and $v_w$=750$\,$km$\,$s$^{-1}$.
This wind
would be capable of driving a shock of speed $v_s$ into
a medium of number density $n_H$ with
$n_Hv_s^2\,\approx\,$1.7$\,\times\,$10$^7$$\times$$(d[''])^{-2}$cm$^{-3}$km$^2$s$^{-2}$,
i.e. $v_s$=160 (800) kms$^{-1}$ at a distance
$d$=5'' (1''), assuming $n_H$=26$\,$cm$^{-3}$. Note that this is an upper
limit for $v_s$, since we have assumed that the ISM contains only atomic hydrogen.
These shock speeds are comparable to the proper motion values of the two sources
and it is unlikely that the resulting bow shock orientations would be along the same line.
Therefore it is clear that a single star cannot be responsible for X3 and X7.
On the other hand, a combined isotropic wind of all the mass-losing stars at the
GC ($\dot{M}_w$=10$^{-3}$\solm$\,$yr$^{-1}$ and $v_w$=750$\,$km$\,$s$^{-1}$), would
result in
$n_Hv_s^2\,\approx\,$1.7$\,\times\,$10$^9$$\times$$(d[''])^{-2}$cm$^{-3}$km$^2$s$^{-2}$, i.e.
$v_s$=1600 (8000)$\,$kms$^{-1}$ at a distance
$d$=5'' (1''). Interestingly, these results
match the above estimate of X3 and X7 being CSPNe or [WC]-stars.
Also, both types of stars are very short-lived, which nicely explains
the lack of other objects of the same morphology in this region.

The number of PNe expected to reside within the central parsec can be estimated as follows:
\begin{equation}
N_{PN} = \sum_{m<8{M_{\odot}}} N_{*}(m)\frac{\tau_{PN}}{\tau(m)},
\end{equation}
where the summation goes over all stars with masses below 8$\,$\solm. We set the lower
mass limit at about 1$\,$\solm, since the stars of lower masses have lifetimes comparable to, or longer than
the Hubble time, and could not yet reach the PN phase. 
\boldmath{$\tau(m)$} is a typical mass-dependent stellar lifetime, and $\tau_{PN}$ is lifetime
of the fast wind of the CSPNe. The latter is also mass-dependent, and ranges from $<$100$\,$yr for 
a 8$\,$\solm~star to $\sim\,$3$\times$10$^4$ yr for a 1$\,$\solm~(see e.g. \citealt{villaver02}).
$N_{*}(m)$ stands for the number of stars of the mass $m$ currently present in the GC population.
\citet{figer04} present the K-band luminosity function (LF) resulting from
deep observations of the several HST/NICMOS fields in the GC region.
To estimate the stellar counts at the faint end of the LF, we fit a power law to the number counts in the
range K$\approx$16.5--19.5, where the data are reasonably complete. We also attempt to correct
for the completeness using the data from \citet{figer99}. By fitting the LF in this range we avoid
the red clump population, but also the population of massive bright stars that dominate the
LF in the central parsec. For scaling, we use the observed (completeness corrected) number counts in the central
parsec, at K$\,$=$\,$17 \citep{schoedel07}, while accounting for the different magnitude
 bin sizes in \citet{schoedel07} and \citet{figer04}. 
This results in probability of $\sim$40$\%$
to find one PNe within this region. 
We observe two sources in a much smaller volume, 
which makes it less likely that they are both of the same short-living type.
This speaks in favor of X3 actually being a dust feature. 

The alignment of the two features is still not explained. In case of an
isotropic wind arising from the mass-losing stars, one would expect
a more random distribution of such sources around the center.
Curiously, the two sources are arranged in the exact direction
in which the mini-cavity is projected onto the sky.
If we do not think of this arrangement as a chance configuration, this might indicate that
$(i)$ all three features (X3, X7 and the mini-cavity) are produced by the same
event, and $(ii)$ there is a preferential direction in which the mass
is expelled at the GC. The possibility of a collimated outflow was already
discussed by \citet{muzic07}. This outflow could also account for narrow
dust filaments of the Northern Arm of the mini-spiral, as well
as the H$_2$-bright lobes of the circum-nuclear disc (CND). As
the authors argue, the outflow could be linked to the
plane of the mass-losing stars in the way that the matter
provided by stars and not accreted onto Sgr$\,$A* is expelled
perpendicular to the plane. Having an opening angle of about 30$^{\circ}$, this
outflow could account for the mini-cavity, X3 and X7 at the same time.
In this case X3 and X7 should be located not too far away from the plane
containing Sgr$\,$A*, which is already suggested by the high inclination (i$\approx$90$^{\circ}$)
of the two bow shocks to the line of sight resulting from our modeling.

\section{Summary and conclusions}
\label{summary}
We have presented L'-band observations of the two
cometary-shaped sources in the vicinity of Sgr$\,$A*, named X3 and X7.
The symmetry axes of the two sources are aligned within 5$^{\circ}$ in the
plane of the sky
and the tips of their bow-shocks point towards Sgr$\,$A*. Our measurements show that
the proper motion vectors of both features are pointing in directions more than
45$^{\circ}$ away from the line that connects them with Sgr$\,$A*.
Proper motion velocities are high, of the order of several 100$\,$km$\,$s$^{-1}$.
This misalignment of the
 bow-shock
symmetry axes and their proper motion vectors, together with
high proper motions, suggest that the bow-shocks must
be produced by an interaction with some external strong wind, possibly coming from
Sgr$\,$A*, or stars in its vicinity. We have developed a bow-shock model
in order to fit the observed morphology and constrain the source of the external wind.
The stellar types of the two stars are not known. 
Moreover, one of the features is likely not a star, but just a dust structure.
It might be located
at the edge of the mini-cavity, and shaped by the same wind that produces the X7 bow-shock.
We discuss the nature of the external wind
and show that
neither one of the features can arise via 
interaction with an external wind originating from a single, mass-losing star.
Instead, the observed properties of the bow-shocks provide evidence for interaction with a fast and strong wind
produced probably by an ensemble of mass-losing sources. Alternatively, a possible source
of the wind could be Sgr$\,$A*.
Shock velocities that can result from such a combined outflow over a distance
assumed for the two features X3 and X7,
match the velocities required to produce the bow shocks
of stars in the late evolution stages of CSPNe or [WC]-stars.
Short lifetimes of such stars can explain the
lack of other similar cometary sources in the central parsec.
We discuss our results in the light of the partially-collimated outflow
already proposed in \citet{muzic07} and argue that
such an outflow, arising perpendicular to the CWS, can account for X3 and X7, as
well as for the mini-cavity.
The collective wind from the CWS has a scale of $\sim$10 arcsec.
On scales of about an arcsecond or less theoretical studies
predict a radius-dependent accretion flow \citep[e.g.][]{narayan98, yuan03}.
Within this region the flow of a major portion of the material  
originally bound for accretion onto Sgr$\,$A* is inverted and the  
material is expelled again towards larger radii.
The presence of a strong outbound wind
at projected distances from Sgr$\,$A* of only 0.8" (X7) with a mass load  
of 10$^{-3}$\solm$\,$yr$^{-1}$ is in fact in agreement with models that predict a  
highly inefficient accretion onto the central BH due to a strongly  
radius dependent accretion flow.

Knowledge of wind parameters for two bow-shock stars is crucial
for drawing more quantitative conclusions on the nature of the external outflow.
Spectroscopy should therefore be the next step to confirm our
hypothesis.

\acknowledgements{
Part of this work was supported by the \textit{Deutsche Forschungsgemeinschaft} (DFG) via SFB 494.
RS acknowledges the Ram\'on y Cajal programme of the Spanish Ministerio de Ciencia e Innovaci\'on.
MZ was supported for this research through a stipend from
the International Max Planck Research School (IMPRS) for
Astronomy and Astrophysics at the Universities of Bonn
and Cologne.}

\bibliography{ms_bowshockstars_5}
\end{document}